# Site-resolved hyperfine and lattice dynamical properties of a single-crystal Fe$_3$Si

S. M. Dubiel[1*] and J. Żukrowski[2]

[1]AGH University of Science and Technology, Faculty of Physics and Applied Computer Science, al. A. Mickiewicza 30, 30-059 Kraków, Poland, [2]AGH University of Science and Technology, Academic Center for Materials and Nanotechnology, al. A. Mickiewicza 30, 30-059 Kraków, Poland

**Abstract**

A single-crystal sample of a DO$_3$-ordered Fe$_3$Si compound was studied by means of the $^{57}$Fe Mössbauer spectroscopy. Spectra have been recorded in a transmission geometry in the temperature range between 5 and 850 K. They have been analyzed either in terms of five sextets or five hyperfine field distribution curves. Concerning the former three sub spectra have been attributed to Fe atoms occupying the B sites and having 8Fe, 7Fe1Si and 6Fe2Si atoms in the first coordination shell (1NN), respectively, and two have been attributed to Fe atoms residing on the A and C sites and having 4Fe4Si and 3Fe5Si atoms in the 1NN, respectively. Based on the temperature dependence of the center shifts values the Debye temperature, $T_D$, have been determined. $T_D$ was found to be equal to 491 (11) K for the sites A and C, and to 403 (3) K the site B. The smaller value of $T_D$ coincides with the smaller values of the hyperfine field, H. Changes in H caused by one Si atom present in the 1NN have been determined as $\Delta H \approx 13$ kOe for Fe atoms occupying the A and C sites and $\Delta H \approx 54$ kOe for Fe atoms present on the B sites. These changes have been also expressed in terms of the underlying number of polarized s-like electrons, $\Delta N_s$. For the Fe atoms situated on the A and C sites $\Delta N_s=0.03$ was found while for those occupying the B sites $\Delta N_s=0.17$.

[*]Corresponding author: Stanislaw.Dubiel@fis.agh.edu.pl

# 1. Introduction

Fe$_3$Si is well-known example of the DO$_3$ ordered compound. It can be also regarded as a quasi-Heusler compound, X$_2$YZ, with X and Y = Fe and Z = Si. It orders ferromagnetically at ~800-850 K [1-4] and has rather high spin polarization of ~45%, even for thin-films [5]. Consequently, it has been subject of numerous recent experimental e. g. [6-10] and theoretical e. g. [11-15] studies. In the ideal DO$_3$-ordered structure there are four crystal sites designated in Wyckoff coordinates as: A(0,0,0), B(1/4,1/4,1/4), C(1/2,1/2,1/2) and D(3/4,3/4,3/4). Three of these sites viz. A, B and C are occupied by Fe atoms, while Si atoms reside on the D sites. In the ideal DO$_3$ structure the sites A and C are equivalent, and magnetic Fe atoms are present at two crystallographically different sites. Those present on the A and C sites are indicated as Fe(I), and those present on the B sites as Fe(II). They also have different coordination numbers viz. Fe(I) atoms have 4 Fe and 4 Si atoms in the nearest-neighbor shell (1NN), and Fe(II) atoms have 8 Fe atoms in the 1NN [16-18]. Consequently, magnetic moments of the two-types of Fe atoms are significantly different: $\mu_{Fe}(I)$ = 1.35 $\mu_B$ and $\mu_{Fe}(II)$ = 2.2-2.4 $\mu_B$ as determined from polarized neutrons [17,18]. Correspondingly different are hyperfine parameters at RT viz. the magnetic hyperfine field: $H_{Fe}(I)$ = 20 T and $H_{Fe}(II)$ = 31 T , and the isomer shift (relative to $\alpha$-Fe at RT): $IS_{Fe}(I)$=0.08 mm/s and $IS_{Fe}(II)$=0.26 mm/s [6]. Astonishingly no much information can be found in the literature on lattice dynamical properties of the Fe$_3$Si compound, and in particular on the Debye temperature, $T_D$. Recently Liang et al. performed first-principles calculations of phonon and thermodynamic properties of various Fe-Si compounds [19]. In particular, they found that the predicted phonon density of states (PDOS) is very characteristic of different Fe-Si compounds. Based on the calculated PDOS they calculated values of $T_D$, and found that they span over the range of 458-615 K, 458K being the Debye temperature for Fe$_3$Si. The latter agrees well with the one determined from the PDOS measured at RT viz. 444(7) K [21]. However, different $T_D$-values were also reported. For example, $T_D$=495(10) K was obtained from X-ray integrated intensity measurements at 96 and 300 K [22] and $T_D$=501 K at 0K was derived from the $C_{44}$ elastic constant [23]. Much lower $T_D$-values viz. 250-375 K, were determined from XRD-measurements on a powder sample of Fe$_3$Si subjected to different heat treatments [24]. One of the aims of the present study was to experimentally determine the value of $T_D$ using the Mössbauer spectroscopy as a tool. This technique permits to obtain site-resolved results thanks to the site-sensitive spectral parameters. Noteworthy, to our best knowledge, not such results were reported so far. Of particular interest of this study was to see whether or not the value of $T_D$ was different for the two lattice sites occupied by Fe atoms i.e. Fe(I) and Fe(II) which have very different magnetic properties. In case of a strong-enough spin-phonon coupling the values of $T_D$ should be different.

## 2. Experimental

### 2.1. Sample

The alloy of a nominal composition $Fe_{75}Si_{25}$ has been prepared from 3N-Fe and 2N-Si 0.99 by melting the constituents in an induction furnace at $10^{-1}$ Torr Ar atmosphere and casting into a mould of copper. Single crystal has then been prepared by the method of Bridgman in a $Al_2O_3$ crucible at $10^{\sim 2}$ Torr atmosphere. Crystal of the [110] orientation have been cut and homogenized by 12 h treatment at 1050 °C and subsequent cooling to room temperature at rate of 12 - 14 °C/h. The Si content has been determined by microprobe analysis as equal to 26.2 at% [25]. For $^{57}$Fe Mössbauer-effect measurements a ~30 μm thick platelet and ~25 mm diameter was prepared by mechanical and electro polishing.

### 2.2. Measurements

$^{57}$Fe Mössbauer spectra were measured using a standard spectrometer (Wissel GmbH) and a drive working in a sinusoidal mode. The 14.4 keV gamma rays were supplied by a $^{57}$Co/Rh. Its activity permitted recording a spectrum of a good statistical quality within a 2-3 days run. The spectra were recorded in a 1024-channel analyzer within the temperature range of 5-850 K. For that purpose the investigated sample was placed (a) in a Janis Research 850-5 Mössbauer Refrigerator System ($5 \leq T \leq 90$ K), (b) in a standard Janis SVT-200 cryostat ($80 \leq T \leq 295$ K) and (c) in a home-made Mössbauer furnace ($400 \leq T \leq 850$ K). The temperature in the (a) and (b) cases was kept constant to the accuracy better than ±0.1 K., and in the (c) case ±0.2 K. Examples of the measured spectra in the magnetic phase are shown in Fig. 1, while a paramagnetic spectrum is displayed in Fig. 2.

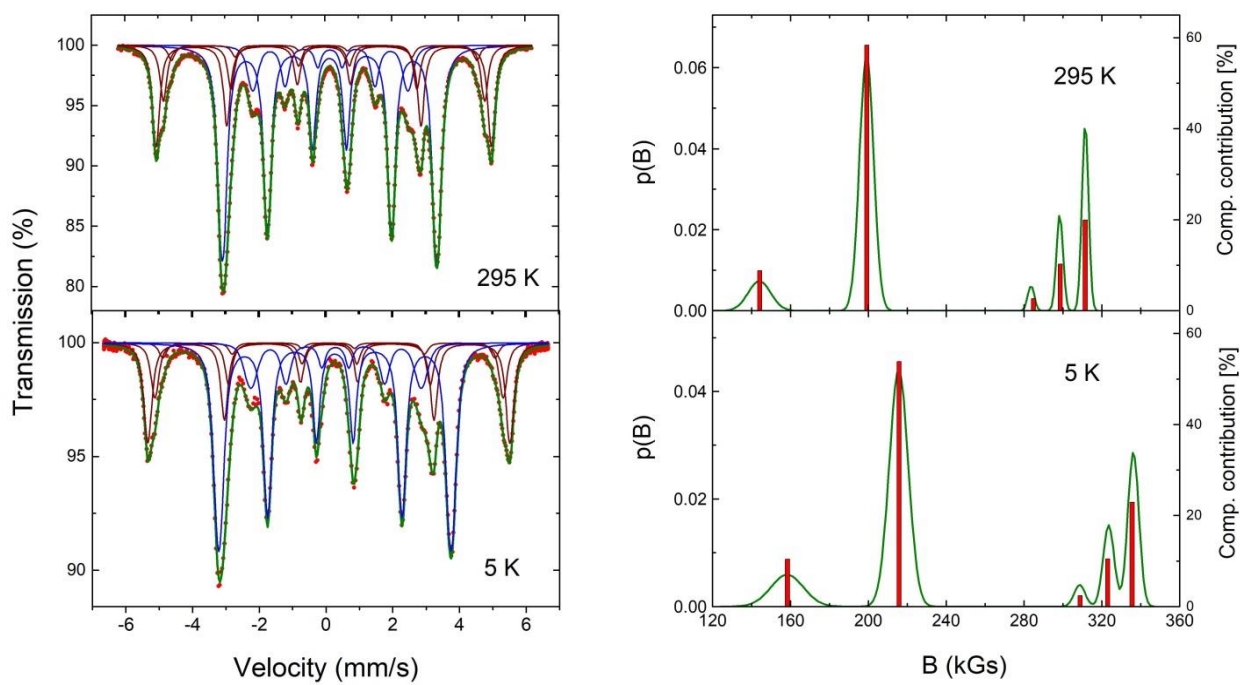

Fig. 1 (Left panel) $^{57}$Fe Mössbauer spectrum recorded at 5 K and at 295 K on the Fe$_3$Si single-crystal sample. The best-fit to the data is indicated by a green line, while five sub spectra by blue lines, (right panel) distribution of the hyperfine field, p(H), illustrating the five different Fe environments. The corresponding values of H$_k$ (k=1, 2, 3, 4, 5) are added as vertical slabs.

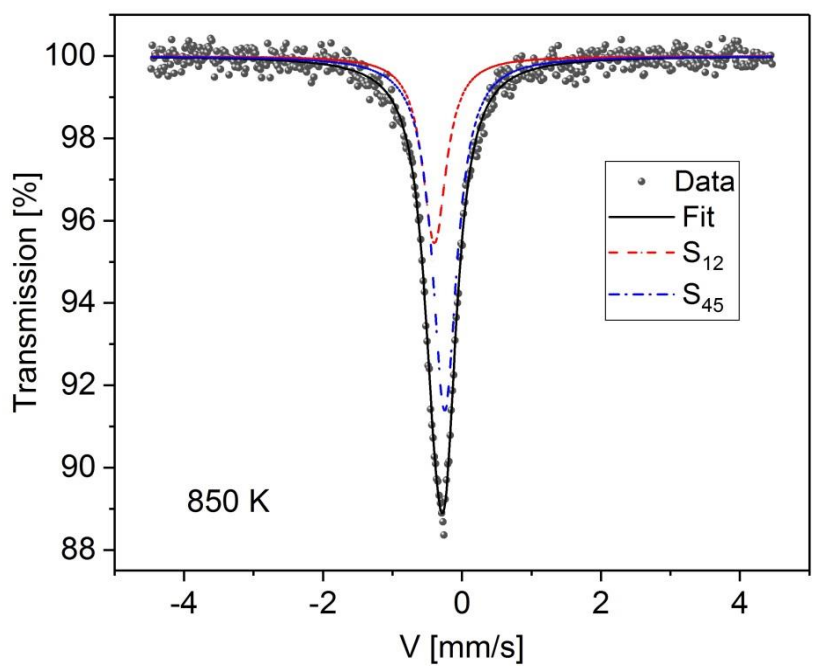

Fig. 2 $^{57}$Fe Mössbauer spectrum recorded at 850 K on the Fe$_3$Si single-crystal sample. Two single-line sub spectra S$_{12}$ and S$_{45}$ are marked.

## 3. Results and discussion

### 3.1. Spectra analysis

As can be seen in Fig. 1, each spectrum consists of more than 12 resonance lines what implicates that Fe atoms have more than 2 different environments. This can have two reasons: (a) lack of stoichiometry and/or (b) chemical disorder. As the composition of our sample is slightly off stoichiometry the reason (a) is most likely responsible for the existence of more than two sextets (Fe environments) necessary to analyze the spectra. It has turned out that the spectra could have been successfully fitted in terms of five components, in form of sextets, corresponding to these environments. For this purpose a transmission integral method and all three hyperfine interactions were taken into account. Thus, for each sub spectrum (sextet) as free parameters were treated: hyperfine field, H$_k$, center shift, CS$_k$, main component of the electric field gradient, V$_{zz}$, line width at half maximum, G$_k$, (it was assumed G$_1$=G$_2$=G$_3$), and relative line intensities within a given sextet (they were common for all 5 sextets). The best-fits spectral parameters obtained based on this procedure are displayed in Table 1. The paramagnetic spectrum shown in Fig. 2 was analyzed in terms of two singlets associated with Fe atoms present on the two types of the lattices sites viz. singlet S$_{12}$, with Fe(I) and S$_{45}$ with Fe(II). The imposed constraints were that their relative abundance was 2:1 and the lines had the same linewidth.

**Table 1.** Best-fit spectral parameters as obtained for the studied sample using the five-sextets fitting method. Values of the center shift are in mm/s (relative to bcc-Fe at RT), those of the hyperfine field in kGs and the line width in mm/s. Values of the spectral area are: A$_1$=20-21%, A$_2$=9-11%, A$_3$=2-3%, A$_4$=54-57% and A$_5$=10-12%. Typical errors of CS$_k$ are ±0.001 mm/s, of the hyperfine field H$_k$ ±0.2 kGs and of the spectral area ±1 %. The average values are marked by < >.

| T [K] | CS$_1$ | CS$_2$ | CS$_3$ | CS$_4$ | CS$_5$ | H$_1$ | H$_2$ | H$_3$ | H$_4$ | H$_5$ | <CS> | <H> |
|---|---|---|---|---|---|---|---|---|---|---|---|---|
| 5.2 | 0.197 | 0.203 | 0.186 | 0.371 | 0.397 | 336 | 323.7 | 308.6 | 215.6 | 158.3 | 0.316 | 247.1 |
| 12.2 | 0.197 | 0.203 | 0.179 | 0.372 | 0.399 | 335.8 | 323.6 | 306.6 | 215.5 | 158.1 | 0.316 | 246.8 |
| 18.6 | 0.200 | 0.204 | 0.187 | 0.373 | 0.401 | 335.8 | 322.8 | 308.2 | 215.5 | 158.5 | 0.318 | 247.2 |
| 25.3 | 0.202 | 0.207 | 0.200 | 0.375 | 0.401 | 335.7 | 323.2 | 308.6 | 215.3 | 157.8 | 0.320 | 246.5 |
| 32.3 | 0.197 | 0.206 | 0.190 | 0.373 | 0.397 | 335.8 | 323.6 | 309.6 | 215.2 | 158.1 | 0.317 | 247.0 |
| 39.2 | 0.195 | 0.201 | 0.184 | 0.373 | 0.396 | 335.8 | 323 | 308.3 | 215.1 | 158.4 | 0.316 | 247.3 |
| 46.2 | 0.196 | 0.199 | 0.183 | 0.371 | 0.397 | 335.5 | 323.1 | 308.9 | 215.1 | 158.2 | 0.316 | 246.7 |

| | | | | | | | | | | | |
|---|---|---|---|---|---|---|---|---|---|---|---|
| 53.3 | 0.194 | 0.197 | 0.195 | 0.369 | 0.388 | 335.1 | 322.7 | 309.2 | 214.9 | 158.1 | 0.313 | 246.4 |
| 60.6 | 0.194 | 0.195 | 0.202 | 0.370 | 0.391 | 334.4 | 321.7 | 307.8 | 214.3 | 157.4 | 0.314 | 246.1 |
| 66.9 | 0.194 | 0.193 | 0.210 | 0.369 | 0.395 | 334.1 | 321.4 | 307.7 | 214.1 | 157.2 | 0.314 | 245.8 |
| 78 | 0.194 | 0.197 | 0.201 | 0.369 | 0.391 | 334.1 | 322.0 | 308.4 | 214.2 | 157.1 | 0.314 | 245.6 |
| 115 | 0.181 | 0.184 | 0.180 | 0.356 | 0.375 | 331.1 | 318.7 | 304.6 | 212.4 | 155.5 | 0.301 | 243.3 |
| 155 | 0.162 | 0.167 | 0.163 | 0.336 | 0.356 | 327.3 | 314.3 | 299.9 | 209.9 | 153.2 | 0.281 | 240.5 |
| 215 | 0.129 | 0.133 | 0.132 | 0.302 | 0.317 | 321.3 | 308.3 | 294.1 | 205.8 | 149.9 | 0.248 | 235.8 |
| 265 | 0.098 | 0.104 | 0.093 | 0.271 | 0.290 | 315.3 | 302.3 | 287.6 | 201.6 | 146.3 | 0.247 | 231.0 |
| 295 | 0.080 | 0.085 | 0.079 | 0.251 | 0.269 | 311.5 | 298.4 | 283.6 | 199.1 | 144.1 | 0.198 | 227.7 |
| 400 | 0.008 | 0.021 | 0.021 | 0.179 | 0.190 | 289.0 | 273.9 | 253.6 | 184.1 | 131.8 | 0.124 | 210.7 |
| 500 | -0.060 | -0.051 | -0.051 | 0.109 | 0.129 | 268.9 | 254.0 | 238.6 | 170.1 | 121.2 | 0.059 | 194.4 |
| 600 | -0.127 | -0.111 | -0.111 | 0.034 | 0.050 | 241.9 | 225.9 | 208.1 | 151.9 | 106.4 | -0.015 | 173.5 |
| 700 | -0.192 | -0.182 | -0.169 | -0.037 | -0.003 | 200.9 | 183.8 | 179.2 | 124.9 | 83.4 | -0.081 | 141.7 |
| 740 | -0.216 | -0.205 | -0.190 | -0.067 | -0.039 | 174.0 | 156.6 | 148.0 | 108.1 | 72.2 | -0.109 | 135.5 |

The five component structure of the measured spectra could be illustratively visualized by analyzing the spectra in terms of a hyperfine field distribution (HFD) method which was based on the code described in Ref. 25. It has turned out that all the spectra could have been satisfactorily fitted in terms of five Gaussian-shaped distributions i.e. the HFD-curves consist of five peaks, as shown in the lower panel of Fig. 1. Each peak, characterized by the hyperfine field, $H_k$ (k=1,2,3,4,5), central shift, $CS_k$, spectral area, $A_k$, and standard deviation, $\sigma_k$. can be ascribed to one site occupied by Fe atoms. The agreement between the spectral parameters obtained with both procedures was very well.

**3.2. Hyperfine Fields**

Temperature dependence of the $H_k$-values (k=1,2,4,5) are displayed in Fig. 3a (The k=3 component has been omitted due to its small, ~2%, contribution).

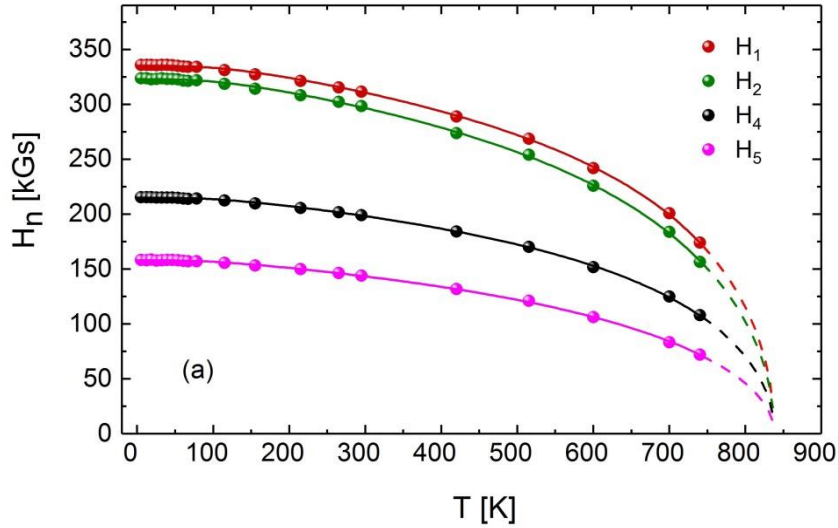

Fig. 3a Hyperfine fields, $H_k$, as obtained for particular Fe environments (k=1, 2, 4, 5) vs. temperature, T. Solid lines represent the best-fits to the data to guide the eye (The same value of the Curie temperature for all components was assumed).

According to the values of $H_k$ and the corresponding ones of the center shift, $CS_k$, (see Fig. 3) two types of Fe environments can be distinguished: Fe (I) and Fe (II). The sub spectra with k=4 and 5 can be associated with the former, hence the Fe atoms occupying the A and C lattice sites, and the sub spectra k=1 and 2 with the latter, hence Fe atoms occupying the B sites. Furthermore, the sextet with k=4 can be attributed to the Fe atoms having 4 Fe and 4 Si atoms in the 1NN, and the sextet with k=5 to those Fe atoms that have 3 Fe and 5 Si as the nearest-neighbor environment. On the other hand, the sextet with k=1, 2, and 3 can be attributed to Fe atoms having 8 Fe atoms in the 1NN (k=1), 7 Fe and 1 Si atoms in the 1 NN (k=2) and 6 Fe and 2 Si atoms in the 1 NN (k=3), respectively. This designation is in line with previous interpretations e. g. [2,6]. The data displayed in Table 1 give evidence that the effect of one Si atom being the nearest-neighbor on the Fe-site hyperfine filed significantly depend on the lattice site. Thus, for the Fe atoms occupying the B sites a decrease of the field amounts to ~13 kOe and for the ones sitting on the A and C sites the decrease is equal to ~54 kOe. The Fe site with the smallest value of the hyperfine field, $H_5$, behaves differently as shown in Fig. 2b, where a reduced values of the hf. fields, $H_i/H_{io}$, are plotted ($H_{io}$ are the values of $H_i$ at T=5 K).

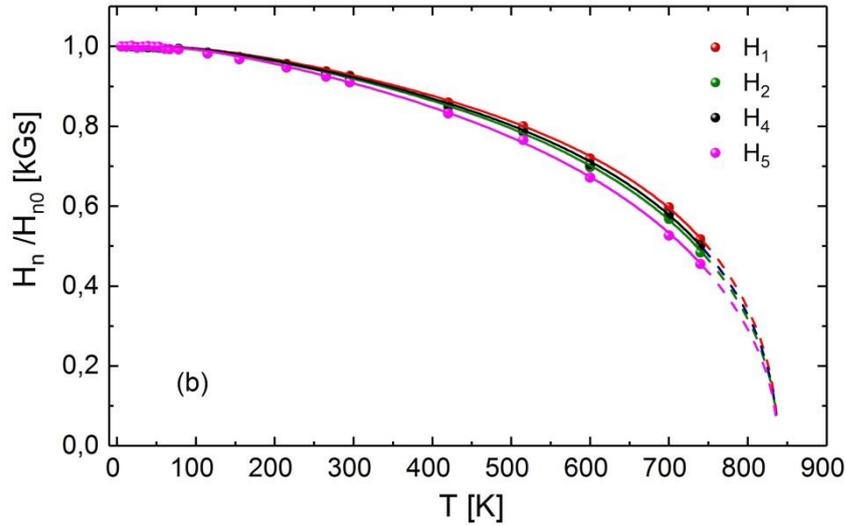

Fig. 3b Reduced values of the hf. field, $H_n/H_{n0}$, as a function of temperature, T. Note the anomalous behavior of $H_5$.

**3.3. Center Shifts**

The temperature dependences of the center shifts are presented in Fig. 4a. Here we can clearly see that $CS_1 \approx CS_2$ ($CS_3$-values have not be included in this figure due to a small, ~2%, contribution of the corresponding sub spectrum), and that $CS_4 \approx CS_5$. The latter values of the center shifts associated with the lattice site A and C are significantly larger than the ones ascribed to Fe atoms occupying the site B ($CS_1$ and $CS_2$). This means that the underlying charge density at Fe atoms at the sites B is correspondingly higher. The issue is discussed in more quantitative way in the paragraph 3.4. The temperature dependence of $CS_k$ can be used to determine the Debye temperature, $T_D$. Following the Debye model approximation, there is the following relationship between CS and $T_D$:

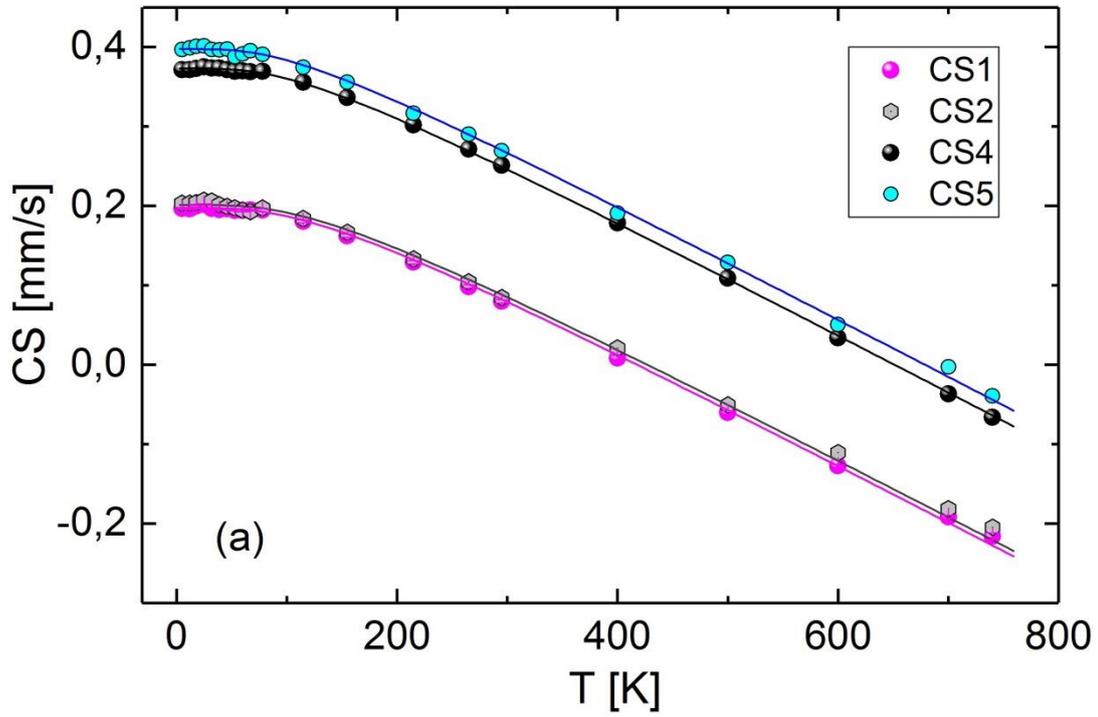

Fig. 4a. Temperature dependence of the central shift for particular environments, $CS_k$ (k=1,2,4,5). The solid lines represent the best-fits of the data to equation (1).

$$CS(T) = IS(0) - \frac{3k_B T}{2mc}\left[\frac{3T_D}{8T} + \left(\frac{T}{T_D}\right)^3 \int_0^{T_D/T} \frac{x^3}{e^x - 1} dx\right] \quad (1)$$

Where IS(0) stays for the isomer shift (temperature independent), $k_B$ is the Boltzmann constant, $m$ is a mass of $^{57}$Fe atoms.

The $CS_k$ values displayed in Tables 1 and 2 were fitted to Eq. (1). Because $CS_1 \approx CS_2$ and $CS_4 \approx CS_5$ for determining corresponding values of $T_D$, the weighted average values $<CS>_{12}$ and $<CS>_{45}$ were fitted to eq. (1). Corresponding data, curves and obtained values of $T_D$ are presented in Fig. 4b.

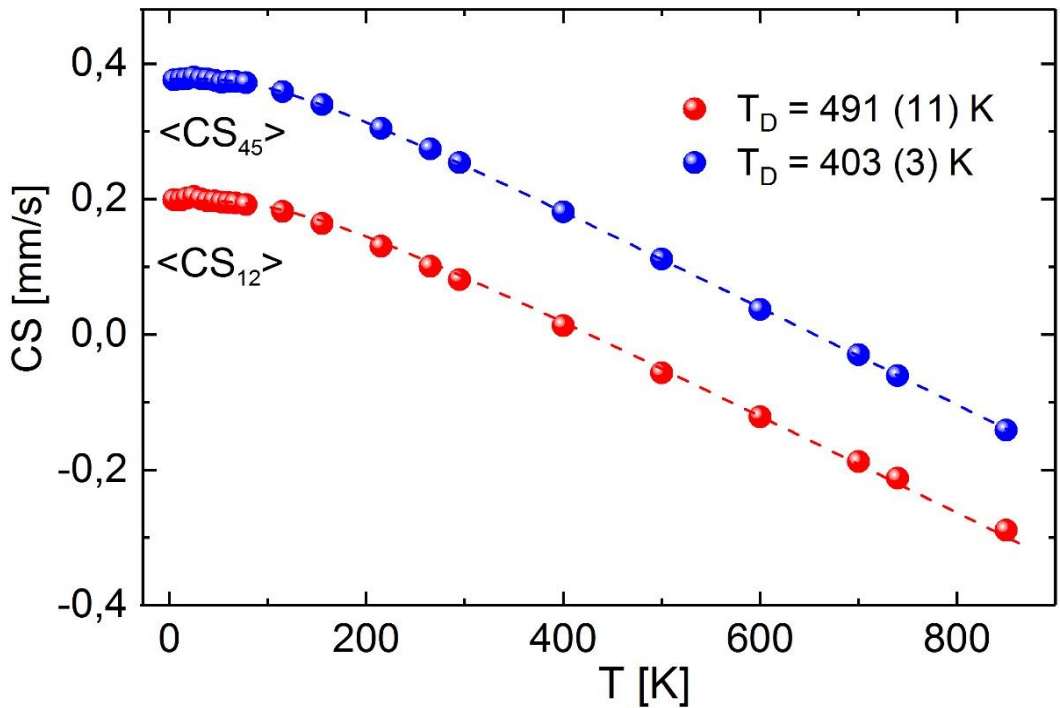

Fig. 4b. Temperature dependence of the average central shifts, <CS$_{12}$> and <CS$_{45}$>. The solid lines represent the best-fits of the data to equation (1). Values of the derived Debye temperature, T$_D$ are indicated.

Clearly, the T$_D$-value characteristic of the sites A and C is significantly different than the one characteristic of the site B. This means that the vibrations of Fe atoms occupying crystallographically different lattice sites in the Fe3Si compound are different. This two types of the sites have different magnetic properties viz. magnetic moments and hyperfine fields. The smaller value of T$_D$ coincides with the smaller values of these magnetic properties. This correlation can be likely regarded as an argument in favor of a spin-phonon coupling. Noteworthy, The average value of T$_D$ agrees pretty well with those reported in the literature and obtained with different methods [21-23].

**3.4. Electronic Structure**

Behind Si-induced changes in the hyperfine field and in the center shift, at a given temperature, there are underlying changes in the spin- and charge-densities at nuclei of Fe atoms. As only s-like electrons have non-vanishing density at the nuclei the changes determined from the corresponding spectral parameters depict effective densities of the s-like electrons. These changes can be expressed in terms of a number of the s-like electrons based on correlations between changes in the hyperfine fields and the corresponding changes in the isomer shifts as outlined elsewhere for disordered binary Fe-rich alloys [26]. As can be seen in Fig. 5, for all sub spectra there are nice correlations between the center shift and the related hyperfine field. Consequently, there is also such correlation between the average quantities i.e. <CS> and <H>. All these correlations are fairly well linear – see examples in Fig. 6a and 6b. At a given temperature, a change in CS, $\Delta$CS, is caused by a change in the isomer shift (the first term in equal. (1)), $\Delta$IS, hence in an effective change of the number of s-like electrons, $\Delta N_s$. Similarly, behind a change of the hyperfine field, $\Delta$H, there is a change of the effectively polarized s-like electrons, $\Delta N^{\uparrow\downarrow}_s$. The simplest explanation of the linear correlation between CS and H is that only electrons of one sub band i.e. spin-up ($\uparrow$) or spin-down ($\downarrow$) are changed. In this case $\Delta N_s = \Delta N^{\uparrow}_s$ or $\Delta N^{\downarrow}_s$. As the presence of Si atoms in the 1NN causes a decrease of the Fe-site charge-density (increase of the isomer shift) and also a decrease of the hyperfine field, which for the Fe-X alloys is negative, it follows that the spin-up s-like electrons flow from Fe to Si atoms. In other words, $\Delta N_s = \Delta N^{\uparrow}_s$. In the studied sample there are two different Fe-sites, Fe(I) and Fe(II). For the former $\Delta$H(I)=53.5 kOe at 5K and 55.2 kOe at 295K, and for the latter $\Delta$H(II)=13.3 kOe both at 5 K and at 295K. To calculate from this figures the corresponding values of $\Delta N_s$, we use the correlations shown in Fig. 5 and the relationship between IS and $N_s$ for Fe-alloys [27]. From the former we obtain the following slopes: $\Delta H_1/\Delta IS_1$=227 kOe/mm/s, $\Delta H_2/\Delta IS_2$=222 kOe/mm/s, $\Delta H_4/\Delta IS_4$=135 kOe/mm/s, $\Delta H_5/\Delta IS_5$=204 kOe/mm/s and $\Delta$<H>/$\Delta$<IS>=189 kOe/mm/s.

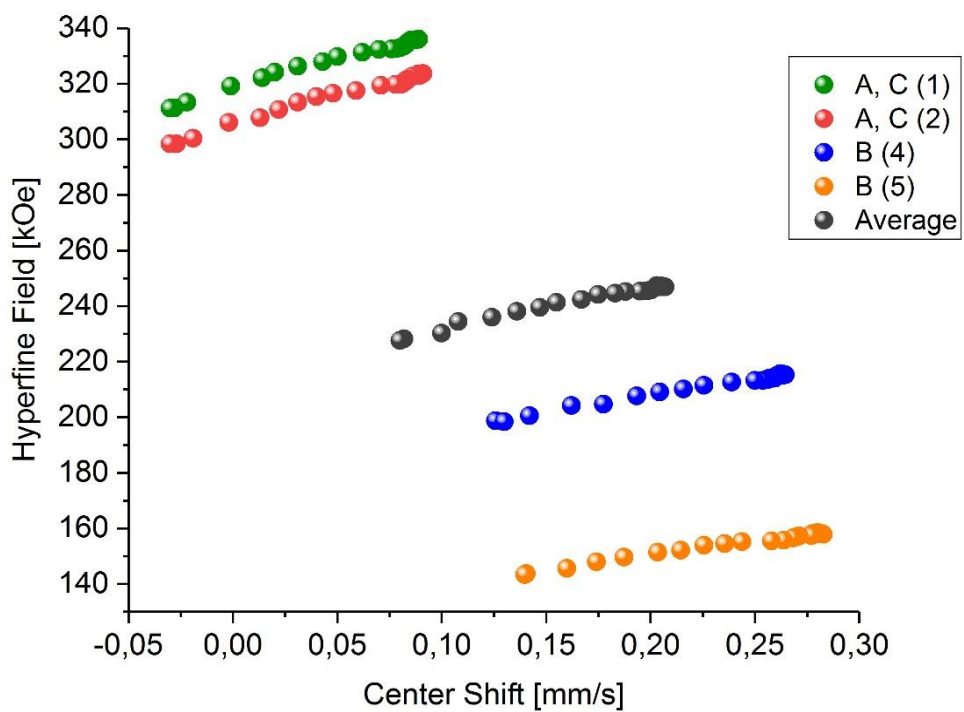

Fig. 5 Relationship between the hyperfine fields and the corresponding center shifts obtained for the environments 1 and 2 of Fe atoms occupying the sites A and C, and for the environments 4 and 5 of Fe atoms occupying the site B.

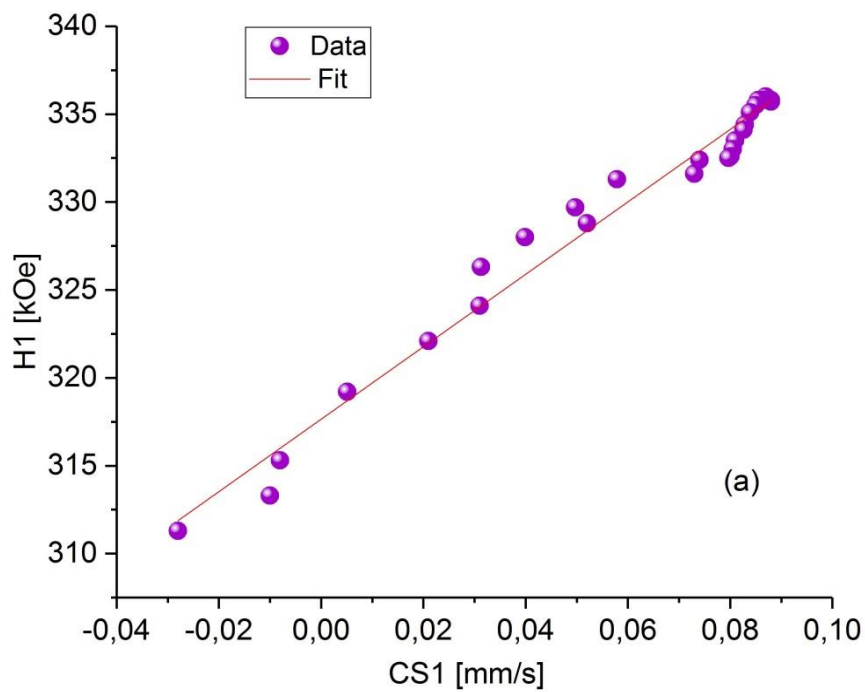

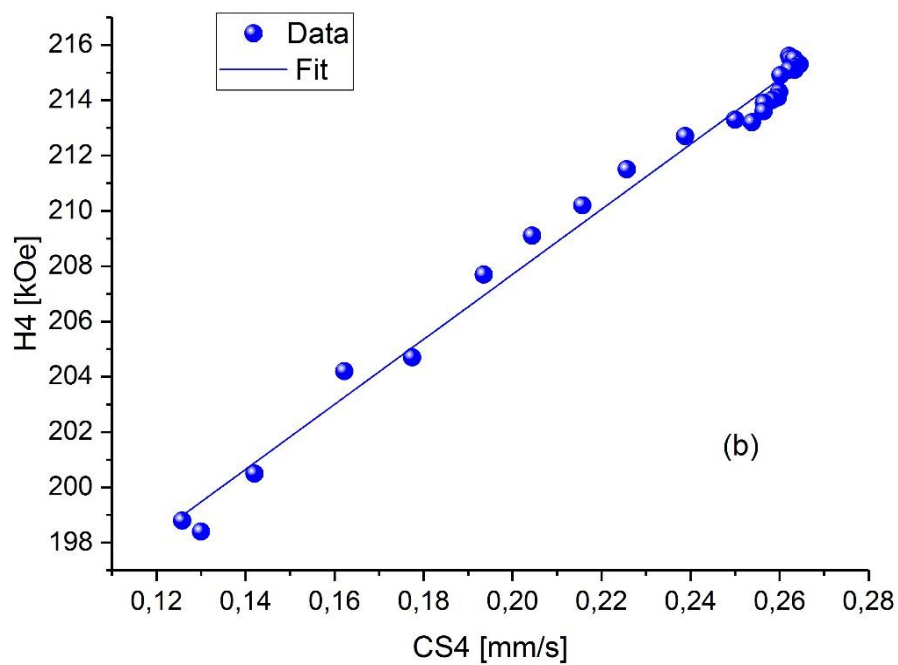

Fig. 6 Relationship between the hyperfine, H, and the corresponding center shift, CS, for Fe atoms on sites: (a) (A,C) in the environment 1, and (b) B in the environment 5. The solid line stands for the best linear fit to the data.

The IS-$N_s$ relationship given in [27] yields $\Delta IS/\Delta N_s \approx 2$ mm/s per s-like electron. Combining both quantities we arrive at the following values of the hyperfine coupling constants, $\Delta H/\Delta N_s$: $\alpha_1=\Delta H_1/\Delta N_s=412$ kOe per 1 s-electron, $\alpha_2=\Delta H_2/\Delta N_s=422$ kOe per 1 s-electron, $\alpha_4=\Delta H_4/\Delta N_s=437$ kOe per 1 s-electron, $\alpha_5=\Delta H_5/\Delta N_s=238$ kOe per 1 s-electron and $<\alpha>=\Delta<H>/\Delta N_s=358$ kOe per 1 s-electron. From these figures we get the corresponding values of $\Delta N_s(k)=\Delta H_k/\alpha_k$, namely $\Delta N_s(1)=\Delta N_s(2)=0.03$ for Fe(II) atoms and $\Delta N_s=0.17$ for Fe(I) atoms (in the latter case the average value of $\alpha_4$ and $\alpha_5$ was used). Noteworthy, in the disordered Fe-rich Fe-Si alloys 1 Si atom present in the 1NN shell decreases the Fe-site charge(spin)-density by 0.025 s-like electron [28], hence similarly like in the present case for the A and C sites.

**4. Conclusions**

Results obtained in this study permitted the following conclusions to be drawn:

1. Fe atoms are present at different crystallographic sites: A and C, called Fe(II) and B, called Fe(I).

2. There are three different environments of Fe(II) atoms attributed to three different atomic configurations viz. with 8 Fe, 7 Fe1 Si, 6 Fe2 Si atoms in the 1NN shell.

3. There are two different environments of Fe(I) atoms attributed to two different atomic configurations viz. with 4Fe4Si and 3Fe5Si atoms in the 1NN shell.

4. Values of the hyperfine field, H, and those of the center shift, CS, could be uniquely associated with each of the five environments.

5. The H-values characteristic of the five environments have significantly different values.

6. The CS-values are characteristic of the lattice site i.e. $CS_1 \approx CS_2 \approx CS_3$ for Fe(II) and $CS_4 \approx CS_5$ for Fe(I).

7. Values of the Debye temperature, $T_D$, are characteristic of the lattice site viz. $T_D(II)=491$ (11) K and $T_D(I)=403$ (3) K.

8. Change of the hyperfine field by 1 Si atom, $\Delta H$, present in the 1NN shell is equal to ~13 kOe for the Fe(II) atoms and to ~54 kOe for the Fe(I) atoms.

9. The ΔH-values have been rescaled into the underlying changes in the number of effectively polarized s-like electrons, $\Delta N_s$, yielding 0.03 for Fe(I) and 0.17 for Fe(II).

**Acknowledgements**

This work was financed by the Faculty of Physics and Applied Computer Science AGH UST and ACMIN AGH UST statutory tasks within subsidy of Ministry of Science and Higher Education, Warszawa. The investigated sample was provided by G. Inden.